# Indium selenides for next-generation low-power computing devices

Seunguk Song[1,2,3†], Michael Altvater[4,5], Wonchan Lee[1,2], Hyeon Suk Shin[2,6], Nicholas Glavin[4,†], and Deep Jariwala[3,†]

[1]*Department of Energy Science, Sungkyunkwan University (SKKU), Suwon, Republic of Korea*
[2]*Center for 2D Quantum Heterostructures (2DQH), Institute for Basic Science (IBS), Suwon, Republic of Korea*
[3]*Department of Electrical and System Engineering, University of Pennsylvania, Philadelphia, Pennsylvania , United States of America*
[4]*Air Force Research Laboratory, Materials and Manufacturing Directorate, Wright-Patterson Air Force Base, Ohio , United States of America*
[5]*AV Inc., 4401 Dayton Xenia Rd. Dayton, Ohio , United States of America*
[6]*Department of Chemistry, Sungkyunkwan University (SKKU), Suwon, Republic of Korea*

[†]email: seunguk@skku.edu, nicholas.glavin.1@afrl.af.mil, and dmj@seas.upenn.edu

## Abstract

As silicon-based computing approaches fundamental physical limits in energy efficiency, speed, and density, the search for complementary materials beyond silicon-based technology has intensified. In this Perspective, we examine van der Waals indium selenides—particularly InSe and $In_2Se_3$—as promising candidates for next-generation low-power electronics. Indium selenides exhibit exceptional electron mobility exceeding 1,000 $cm^2V^{-1}s^{-1}$, high thermal velocity of $>2\times10^7$ cm $s^{-1}$, thickness-tunable bandgaps of 0.97-2.5 eV, and unique phase-dependent ferroelectric properties, enabling both high-performance logic and non-volatile memory functions within a single material system. This perspective critically evaluates the materials properties, fabrication challenges, and device applications of indium selenides, examining their potential to surpass silicon in ultra-scaled transistors through ballistic transport while simultaneously offering ferroelectric memory capabilities impossible in conventional semiconductors. We analyze breakthroughs in ballistic InSe transistors, tunnel field-effect transistors, and $In_2Se_3$-based ferroelectric devices for information storage and identify key research priorities for addressing persistent challenges in scalable synthesis, phase control, and oxidation prevention. By bridging fundamental materials science with practical device engineering, we provide a roadmap for translating the exceptional properties of indium selenides into commercially viable, low-power computing technologies that can overcome the limitations of silicon while enabling novel computing architectures.

## Introduction

The miniaturization of silicon-based microelectronics faces concurrent challenges of limited device scalability and increasing energy consumption in advanced processing nodes, necessitating the exploration of novel materials solutions for future low-power computing hardware. The two-dimensional (2D) van der Waals indium selenides—InSe and $In_2Se_3$—have emerged as particularly promising candidates, offering advantages that potentially exceed those of both silicon and other widely studied 2D materials such as transition metal dichalcogenides (TMDs) of Mo and W[1-5]. Their atomically thin form factors yield exceptional electrostatic gate control, mitigating short-channel effects that plague deeply scaled silicon transistors[1,4,5]. Second, InSe exhibits extraordinarily high electron mobility, with experimental values exceeding 1,000 $cm^2V^{-1}s^{-1}$ at room temperature[6-8], owing to an unusually low effective electron mass of ~$0.12m_0$). This high mobility supports ballistic transport in nanoscale devices[6,9,10]. Moreover, the binary indium-selenium system offers a rich phase landscape with diverse structural configurations and electronic properties, allowing for precise engineering of device characteristics through phase and thickness control[11-13].

Perhaps most significantly, certain phases of both $In_2Se_3$ and InSe exhibit robust ferroelectric properties persisting down to the monolayer limit, combining semiconducting behavior with non-volatile memory functionality in a single material platform[14-17]. This integration of logic and memory capabilities within one material system offers an elegant path toward overcoming the energy-intensive data movement between separate logic and memory components in conventional von Neumann architectures[15,18,19].

The materials diversity and functional richness of indium selenides have enabled electronic devices specifically designed for logic and memory applications. For logic devices, substantial progress has been demonstrated with InSe ballistic transistors achieving a record-high transconductance value of ~6 mS and current density of 1.2 mA $\mu m^{-1}$ at $V_{DD}$ = 0.5 V, making them promising candidates for sub-1 nm node integration[6,9]. Additionally, InSe-based tunnel field-effect transistors (TFETs) show potential for ultra-low power switching through sub-thermionic subthreshold characteristics[20]. For memory, $In_2Se_3$ ferroelectric semiconductor field-effect transistors (FeSFETs) leverage inherent ferroelectric properties for non-volatile data storage[15,18], while ferroelectric semiconductor junctions offer innovative approaches to resistance-based memory with high on/off ratios[19].

In this perspective, we critically evaluate advances in indium selenide materials and devices, assessing their potential as cornerstones for beyond-silicon electronics, focusing on low-power computing applications. We analyze the diverse phases and electronic structures and discuss key challenges in scalable synthesis and device integration. By bridging materials science fundamentals with device engineering considerations, this perspective aims to provide a comprehensive assessment of indium selenides' prospects for transforming future micro and nanoelectronics.

## Indium Selenides across Diverse Phases

Layered 2D InSe (**Fig. 1a**, left) and $In_2Se_3$ (**Fig. 1a**, right) have emerged as interesting avenues for materials and device-level exploration due to their superlative electronic and optical properties, relative thermodynamic stability at room temperature, and the ability to grow large crystals or thin films that are easily processable[11-13]. The larger indium selenide family of compounds exists in multiple different polymorphs and polytypes, including layered (2D) InSe, $In_2Se_3$, $In_3Se_4$ (Ref.[21]), and $In_4Se_3$ (Ref.[22]) (**Fig. 1a**, center). This diversity broadens the design landscape but also introduces new processing challenges.

### *Polymorphism and electronic structures of InSe*

Indium (II) selenide (InSe) is a layered material wherein each layer is four atoms thick with a covalently bonded, honeycomb, in-plane lattice with $D_{3h}$ symmetry. Individual layers are van der Waals bonded in at least three distinct polytypes related by interlayer translations and 180° with respect to the 120° crystal symmetry. InSe has different stacking orders and space groups (**Fig. 1a**, left and **Supplementary Table 1)**. Density functional theory has been used to calculate the electronic band structures of each InSe phase[23-25] and, in the cases of β-InSe and γ-InSe, corroborated with angle-resolved photoemission spectroscopy measurements[26,27]. As constituent layers share the same structure, the in-plane electronic properties of different stacking orders have only subtle differences arising from the orbital overlap between layers. All of these phases have similar band gaps, as β-InSe exhibits a direct band gap of 1.2 eV[28] about the Γ point, ε-InSe has an indirect gap of 1.4 eV[29] near the Γ point, and γ-InSe is a direct gap semiconductor with a gap around 1.3 eV[30,31] at the B-point in momentum space (**Supplementary Table 1**). Their low energy dispersions are similar to each other, with the conduction band showing a steeper dispersion and a lower effective mass of 0.12 $m_0$, compared to the valence band whose effective mass is ~ 0.40 $m_0$. This steep dispersion contributes to high thermal velocity and low effective electron mass, which are crucial parameters for ultra-scaled transistor operation[10,32,33]. The thermal velocity of 2D electrons is dependent only on the temperature ($T$) and effective electron mass ($m_e$) as[10,32,33]:

$$v_{th} = \sqrt{\frac{\pi k_B T}{2m_e}}$$

Calculations of the effective masses of charge carriers in bulk InSe reveal very small effective electron masses of 0.14 $m_0$ and 0.08 $m_0$ for in-plane and out-of-plane conduction, respectively; much smaller than other 2D semiconductors such as $MoS_2$ and $WSe_2$ and comparable to 3D semiconductors like GaAs and Si (**Fig. 1b**). Although γ-InSe has received the most attention, all three of the InSe phases are expected to demonstrate high in-plane electron mobility due to their low effective masses and nearly identical low energy dispersions.

In addition to electronic transport, the symmetry of each stacking configuration gives rise to distinct physical functionalities, such as nonlinear optical activity and ferroelectricity. For example, the lack of inversion symmetry in the ε and γ phases allows for non-linear optical effects such as second harmonic generation[34] and piezo-phototronic effect[35] compared to β-InSe. Additionally, the non-centrosymmetry allows additional vibrational modes to be Raman active in the γ and ε polytypes which can couple with other dipolar excitations[36].

Beyond these symmetry-enabled effects, structure-driven ferroelectricity in InSe arises from interlayer stacking dynamics. Multilayer van der Waals InSe exhibits ferroelectric polarization through a mechanism known as sliding ferroelectricity, where the stacking sequence of atomic layers dictates the electrical polarization[16,37] (**Box 1**). Specific stacking arrangements induce charge transfer between layers, resulting in an out-of-plane spontaneous polarization. When these atomic layers slide relative to each other, altering their stacking sequence, the associated charge transfer reverses, thereby flipping the spontaneous polarization. This mechanism is fundamentally distinct from classical displacive ferroelectric and emerges through substitutional doping of Y[16,37] and layer-number engineering[38]. This stacking-dependent polarization makes InSe particularly attractive for ferroelectric applications.

Another key aspect of InSe polymorphism is its strong thickness dependence on the band structure, governed by quantum confinement effects, that allows for the tunability of their electronic and optical properties with material thickness. For example, the optical band gaps of TMDs such as $MoS_2$ and $WS_2$ increase from their bulk values of ~1.5 and ~1.4 eV (indirect) to ~1.9 and ~2.1 eV (direct) in their monolayer limits, respectively[39,40]. Similarly, the thickness dependence of the band gaps of γ-InSe and β-InSe have been measured to be ~1.3 eV (direct) in the bulk and increases to ~2.4 eV (indirect) in the monolayer limit[6,28,41] (**Fig. 1c** and **Box 2**).We also note that, in $In_2Se_3$ system, α-$In_2Se_3$ increases from ~1.3 eV (direct) in bulk to ~1.5 eV (direct) in a monolayer[42], and β- $In_2Se_3$ goes from ~1.43 eV (direct) in bulk to greater than ~1.58 eV (direct) in the bilayer[43]. Unlike Mo and W dichalcogenides, which experience an abrupt transition from direct to indirect band gap going from monolayer to bilayer, these phases of InSe and $In_2Se_3$ gradually transition over thicknesses of ~10 nm, allowing for a high degree of band structure optimization (**Fig. 1c**).

The origin of the direct-to-indirect gap crossover in the InSe phases has been observed experimentally and studied theoretically. Quantum confinement effects cause different valence bands to shift in energy at different rates with decreasing thickness[25,28,44]. As layer number decreases from bulk, the valence band peaks at the Γ or B point get pushed to lower energies, eventually transitioning below the next highest energy valence band between 6-20 layers[25,28]. The valence band at low thicknesses has a maximum along the Γ-K direction, forming a distorted potential. This transition changes the band gap from direct momentum at bulk-like thicknesses to indirect momentum at lower thicknesses (**Box 2**). Due to the reduced dimensionality, many-body effects and long-range forces have a stronger effect on optical and electronic properties. For example, Coulomb interactions and electron-phonon coupling are enhanced in the monolayer limit of InSe, leading to a dramatic increase in excitonic binding energy[25] and dark exciton luminescence[41]. Taken together, the strong dependence of band structures of InSe on layer thickness and stacking order demonstrates how structural and dimensional engineering can be leveraged to tailor its electronic and optical functionalities for emerging device concepts.

### ***Structure and ferroelectricity of $In_2Se_3$***

Polymorphism in van der Waals $In_2Se_3$ is more complicated than in the monoselenide as there are multiple possible structural configurations of each quintuple layer (Se-In-Se-In-Se) in addition to different stacking orders. Thus, the identification and refinement of structural parameters and the resulting electronic properties of these polymorphs remain a central topic of research on phase-dependent phenomena in $In_2Se_3$. The three most commonly reported phases of $In_2Se_3$ are α-$In_2Se_3$, β-$In_2Se_3$, and γ-$In_2Se_3$. In α-$In_2Se_3$, one plane of In atoms is octahedrally coordinated and the other forms tetrahedra with the Se atoms (**Fig. 1a**, right). In β-$In_2Se_3$, both planes of In atoms are octahedrally bonded to Se atoms. The γ phase of $In_2Se_3$ exhibits a defective wurtzite-like nonlayered crystal structure. Because γ-phase $In_2Se_3$ does not have the van der Waals characteristic of 2D layered materials, it is not discussed in detail. Additionally, the α and β phases of $In_2Se_3$ can exhibit different stacking orders. α-$In_2Se_3$ can form both 2H and 3R layered structures with $P6_3/mc$ and $R3m$ space groups, respectively, while β-$In_2Se_3$ can adopt 1T, 2H, or 3R lattices with $P\bar{3}m1$, $P\bar{3}m1$, and $R\bar{3}m$ space groups, respectively[45] (**Supplementary Table 1**). In their monolayer form, β-$In_2Se_3$ polytypes are centrosymmetric, whereas α-$In_2Se_3$ is non-centrosymmetric. This non-centrosymmetry gives rise to inherent electrical polarization even in a monolayer α-$In_2Se_3$[14,46]. Together, these diverse stacking configurations and symmetry variations underpin the distinct electronic and ferroelectric behaviors observed in different $In_2Se_3$ phases.

Ferroelectricity is a unique property of certain materials, characterized by a spontaneous electric polarization that can be reversed by applying an external electric field. In the α-phase $In_2Se_3$, the third-layer Se atoms deviate from their centrosymmetric positions to reach a more stable potential energy state. Upon an external electric field, these Se atoms can shift either downward below the upper In sublayer or upward above the lower In sublayer, depending on the field direction, thereby enabling out-of-plane ferroelectricity (**Fig. 1a**, right). According to theoretical symmetry considerations, the threefold rotational symmetry around the out-of-plane axis in monolayer α-$In_2Se_3$ cancels any possible in-plane polarization, allowing only out-of-plane ferroelectricity[47]. Nevertheless, clear in-plane polarization signals in α-$In_2Se_3$ through piezoresponse force microscopy[15,46,48], second-harmonic generation[49], and opto-electronic transport measurements[49] have been experimentally reported. This discrepancy between theory and experiment suggests that the origin of the observed in-plane polarization remains under debate and requires further investigation. In multilayer α-$In_2Se_3$, the 2H or 3R stacking configuration can further influence its ferroelectric properties[50]. The stacking order modulates interlayer interactions and charge redistributions, leading to variations in the motion of ferroelectric domain walls in both in-plane and out-of-plane directions[50]. Notably, 3R-stacked α-$In_2Se_3$ exhibits in-plane movement of out-of-plane ferroelectric domain walls, resulting in a broader electrical hysteresis, known as the memory window, compared to its 2H-stacked counterpart[50]. The Curie temperature of α-$In_2Se_3$ is reported to be ~700 K[51], which is higher than most of the 2D ferroelectric materials such as ~320 K for van der Waals $CuInP_2S_6$, suggesting its robust operation at higher temperatures (see the comparisons in **Fig. 1d** and **Supplementary Table 2**).

In the case of β-phase $In_2Se_3$, the bulk crystal is centrosymmetric. However, its surface can develop periodic nanostripes that break inversion symmetry, forming a metastable β′ phase (**Fig. 1a**, right). This β' phase is similar to the β phase but with a small displacement of the central Se atoms[52,53]. As this displacement breaks the centrosymmetry of the β phase, the β' phase exhibits in-plane (anti)ferroelectricity, which persists down to the monolayer limit and is stable up to 200 °C in both bulk and thin exfoliated flakes[54,55]. Further, due to an unusual competition between ferroelectric and antiferroelectric ordering, the β' phase often exhibits antiparallel adjacent ferroelectric domains, forming a striped superstructure[52,55,56]. Depending on the relative size and number of alternating domains, the macroscopic crystal can be observed to be antiferroelectric, ferrielectric, or can be poled to become fully ferroelectric[56,57]. Angle-resolved photoemission spectroscopy measurements of the β' phase show an indirect band gap of 0.97 eV in the bulk with an effective electron mass as small as $m_e \sim 0.21\, m_0$[58], while the band gap of monolayer β'-$In_2Se_3$ has been found to be 2.5 eV by scanning tunneling microscope measurements[59]. This gives the β'-$In_2Se_3$ phase a larger range of possible band gap values and thus, more tunability than the β-$In_2Se_3$ phase.

Moreover, small energy differences and energy barriers between different stacking orders and polymorphs can easily lead to the formation of stacking faults and phase impurity[50]. In addition, these polymorphs are capable of undergoing phase transitions driven by temperature, defects, or synthesis parameters, enabling further engineering of the ferroelectric properties within the material system. For instance, heterophase junctions composed of α and β' structures facilitate enhanced non-volatile memory performance through band engineering[12]. Although careful correlation between phase composition and its impact on electronic properties has not yet been demonstrated, proper control and characterization of indium selenide polymorphs and polytypes is expected to be important for producing reliable and reproducible device performance.

## Scalable Thin-Film Deposition

Although InSe and $In_2Se_3$ exhibit attractive material properties, numerous challenges arise in their preparation. The inherent complexity of the In-Se phase diagram complicates the chemical synthesis of materials with the desired stoichiometry (**Fig. 1a**). Unlike 2D TMDs such as $MoS_2$ and 3D III-V compound semiconductors such as GaN, GaAs, and InP, which each exhibit only one stable polymorph, the binary In-Se system features at least four stable phases at room temperature: InSe, $In_2Se_3$, $In_6Se_7$, and $In_4Se_3$. This complexity hinders precise stoichiometric control during chemical synthesis, compromising the phase purity of the resulting crystal. Moreover, InSe is highly unstable under ambient conditions, particularly in the presence of water molecules and oxygen, leading to oxidation. InSe and $In_2Se_3$ spontaneously oxidize rapidly when exposed to air or moisture, following simple reactions:[60]

$$12InSe + 3O_2 \rightarrow 4In_2Se_3 + 2In_2O_3$$

$$2In_2Se_3 + 3O_2 \rightarrow 2In_2O_3 + 6Se$$

This oxidation process results in the formation of amorphous $In_2Se_3$ or elemental Se and converts the crystalline InSe or $In_2Se_3$ into stable $In_2O_3$. Consequently, it is essential to have a thorough understanding of methods for preparing InSe or $In_2Se_3$ with the desired stoichiometry and chemical purity. Furthermore, considering the scalability required for electronic device applications, achieving large-area synthesis of these materials with controlled thickness is equally important.

***Mechanical and chemical exfoliations.***

The preparation of 2D InSe and $In_2Se_3$ nanosheets has employed various top-down techniques, including mechanical and chemical exfoliations from bulk single crystals. Mechanical exfoliation often uses adhesive tapes to produce high-quality, single-crystalline, few-layer flakes from a mother crystal, suitable for fundamental studies. For example, thickness-dependent quantum confinement affecting the band structure[61] and excitonic lasing properties[62] has been investigated in mechanically exfoliated InSe crystals. However, mechanical exfoliation typically yields small flakes of less than 1,000 $\mu m^2$ with limited scalability for commercialization.

Another technique, liquid-phase exfoliation, uses solvents and ultrasonic energy to disperse layered materials into nanosheets. This approach can produce larger quantities of exfoliated material, making it more suitable for applications requiring bulk production.

However, it can introduce defects or impurities, potentially affecting the material's intrinsic properties. A liquid-phase exfoliation using KOH, achieving thicknesses down to 7 nm for InSe was reported in 2018[63]. Here, liquid exfoliation enabled the production of few-layered InSe nanosheets with direct band gaps and large lateral sizes, which are essential for fabricating high-performance photoelectrochemical photodetectors. To prevent oxidation, InSe can be exfoliated using a surfactant-free, deoxygenated low-boiling-point ethanol-water cosolvent system[64]. This approach enables the preparation of high-quality InSe flakes that exhibit exceptional photoresponsivity of ≈$5 \times 10^7$ A $W^{-1}$ in photodetectors.

Another solution-processable method, molecular intercalation, also enables large-scale production of high-quality atom-thick nanosheets with controllable thicknesses[15] (**Fig. 2a**). Electrochemical exfoliation using tetrahexylammonium cations ($THA^+$) in tetraethylammonium bromide in N,N-dimethylformamide (DMF) electrolyte provides a scalable approach to generating atom-thick InSe layers with precise structural control with the expression as below[15]:

$$InSe + xTHA^{+} + xe^{-} \rightarrow (THA^{+})_{x}InSe^{x-}$$

$$xBr^{-} \rightarrow \left(\frac{x}{2}\right) Br_{2} + xe^{-}$$

Upon applied field, $THA^+$ intercalates into the mother crystal bulk-InSe and expands the structure gradually through the intercalation. This eventually results in the production of InSe nanosheets, reducing $Br^-$ ions on the anode. Importantly, under an external electric field, biaxial tension strain induced by the electrochemical process can cause atomic distortion in β-InSe, breaking its inversion symmetry. This distortion enables the emergence of both in-plane and out-of-plane ferroelectricity in InSe[15], indicating the structural and electrical properties are highly dependent on the method. It should be mentioned that a major breakthrough in improving the scalability and high-quality 2D InSe involved developing an oxygen-free electrochemical molecular intercalation process[65]. The resultant InSe monolayers exhibited high purity and uniformity across the 4-inch wafer, while proving its high electrical properties; an average electron mobility of ~90-120 $cm^2V^{-1}s^{-1}$, on/off ratio of ~$10^7$ in fabricated transistors[65].

For $In_2Se_3$, the exfoliation efficiency reaches higher than 83 %, producing sheets up to 30 μm[66]. Furthermore, to date there are no reported cases of $In_2Se_3$ exhibiting ferroelectric properties through liquid-phase processes. Therefore, minimizing the risk of degradation during the chemical exfoliation is crucial for future large-scale production. Note that intercalation of alkali metals such as Li and K has not been demonstrated for InSe and $In_2Se_3$, probably due to the charge transfer-induced degradation issues.

Altogether, advances in top-down exfoliation techniques, ranging from mechanical and liquid-phase to oxygen-free intercalation, have greatly improved the scalability and quality of 2D InSe, while the prevention of chemical degradation remains a key challenge for realizing stable and large-scale production.

***Thin film growth***

Compared with mechanical or chemical exfoliation, thin-film growth of 2D materials offers key advantages for semiconductor industrial applications including thickness and stoichiometry control, enhanced potential for scale-up, gas-phase doping, uniformity across large areas, and the ability to tailor material properties by adjusting growth parameters. However, during chemical vapor-based processes, a Se-rich environment typically favors the formation of $In_2Se_3$, while an In-rich condition increases the possibility of producing InSe[11]. Therefore, to achieve the desired stoichiometric composition, precise control over the fluxes of precursors, especially the chalcogen-to-metal ratio, is essential. Furthermore, the growth temperature and cooling rate also affect the crystalline phases and stacking orders, indicating that multiple growth parameters should be optimized to obtain reliable growth windows.

Metal-organic chemical vapor deposition (MOCVD) emerges as a promising technique for growing phase-pure InSe and $In_2Se_3$ thin film[11,67], particularly in comparison to powder-based thermal chemical vapor deposition (CVD) (**Fig. 2b** and **2c**). Metal-organic sources, such as $In(CH_3)_3$, have lower melting points than metal or metal oxide sources, such as In or $In_2O_3$ (**Supplementary Table 3**). This lower melting point facilitates improved control of the sources, particularly because MOCVD utilizes a bubbler system equipped with electrical pressure controllers and mass flow controllers, enabling precise regulation of the vapor pressure. It is also worth noting that group-III metal precursors, such as In, $In(CH_3)_3$, Ga and $Ga(CH_3)_3$, exhibit lower evaporation or sublimation temperatures than transition metal precursors, such as Mo and W (**Supplementary Table 3**). For instance, $In(CH_3)_3$ has a melting point of 88 °C, and $Ga(CH_3)_3$ is in a liquid state at room temperature, whereas W has an exceptionally high melting point of 3,422 °C. Furthermore, the thermal decomposition temperatures of $In(CH_3)_3$, and $Ga(CH_3)_3$ begin at ~101-150 °C and are lower than those of metal-organic precursors for 2D TMDs at ~250-350 °C for $Mo(CO)_6$ and $W(CO)_6$. This lower temperature requirement for group-III precursors thus provides an advantage for Si back-end-of-line integration at low temperatures of less than 450-550 °C[68].

Compared to InSe, $In_2Se_3$, the In-Ga-Se ternary system remains largely unexplored, yet offers opportunities for material engineering[69-72]. Forming $In_{1-x}Ga_xSe$ alloys could allow for the controlled adjustment of band structures, effective mass, and optical properties beyond phase or thickness engineering alone, using growth techniques such as thermal CVD, MOCVD, chemical vapor transport, and molecular-beam epitaxy (**Fig. 2b-d**). Thermal CVD is the most widely adopted method for growth of materials in the In-Ga-Se ternary system due to its easy accessibility, and it can grow relatively large grain sizes of ~10–100 µm with moderate growth rates of several nm per min by manipulating solid-state precursors (**Fig. 2e** and **2f**). However, sublimation of the sources requires high growth temperatures of 700–900°C (**Fig. 2e**) and precisely controlling the metal-to-chalcogen ratio. Hence, powder-based CVD of stoichiometric InSe remains difficult, with only two related reports available as of 2025[73,74]. Chemical vapor transport, which employs InSe source powder and $NH_4Cl$ as a transport agent within a sealed quartz tube, also enables the growth of few-layer InSe or $In_2Se_3$ at the growth temperature of ~400-450 °C[75]. High-quality $In_2Se_3$ can also be obtained by physical vapor deposition or quasi-equilibrium growth[76] using $In_2Se_3$ powder at high temperatures of up to ~850 °C[43,77,78]. However, the physical vapor deposition of InSe lacks control over In-to-Se ratio. In addition, both chemical vapor transport[75] and physical vapor deposition[43,77,78] typically produce randomly oriented few-layer flakes with smaller grain sizes of less than 20 µm and lack scalability and thickness controllability. In contrast, MOCVD presents a promising approach for scalable and high-quality InSe or $In_2Se_3$ production, offering epitaxial mode, wafer-scale growth, moderate growth temperatures of ~400-500 °C, and thickness controllability[11,67,79]. Molecular-beam epitaxy can similarly achieve large-scale, thickness-controlled InSe or $In_2Se_3$ by adjusting growth temperature and source flux ratio[80,81] (**Fig. 2d**), but has limited throughput for producing large-area thin films and high cost because it requires ultra-high vacuum of $10^{-10}$ torr, which restricts its commercialization when compared with MOCVD (**Fig. 2f**). However, despite advancements in growth techniques, the electrical performance of synthesized indium selenide films lags behind that of mechanically exfoliated (**Fig. 2g**). Specifically, the field-effect mobility of InSe films grown by chemical vapor deposition is typically ~1-10 $cm^2V^{-1}s^{-1}$, much lower than their mechanically exfoliated counterparts of ~$10^3$ $cm^2V^{-1}s^{-1}$ for InSe flakes with the thickness of ~5 nm. This reduced electrical mobility in synthetic films is largely attributed to defects such as undesired oxides, vacancies, grain boundaries, and phase impurities, which hinder charge carrier transport.

Consequently, there is a pressing need for improved growth methods to minimize defects. One promising approach involves the use of flow-modulation MOCVD to optimize the ripening stage, thereby reducing nucleation densities and promoting lateral growth[11,79]. Additionally, unidirectionally oriented epitaxial growth, similar to the epitaxial growth of TMDs single crystals[82], holds potential for producing high-quality thin films. The step-guided epitaxial growth of InSe or $In_2Se_3$ on a vicinal-stepped substrate, such as sapphire, with an off-cut angle has yet to be observed. Complementary to these epitaxial approaches, the solid-liquid-solid growth method reported in 2025 represents a highly promising route toward achieving stoichiometric and highly crystalline InSe[83]. Furthermore, advanced in-situ characterization techniques during growth could enable real-time monitoring of phase formation, potentially allowing for automated feedback systems that dynamically adjust growth parameters to maintain stoichiometric control across entire wafers. The integration of machine learning approaches for optimizing the complex growth parameter space holds promise for rapidly identifying conditions that maximize mobility while minimizing defect density.

Achieving high-quality indium selenide thin films requires careful control of precursor ratios, growth temperature, and cooling conditions to ensure the proper stoichiometry and crystal phase purity for optimal film performance. Further progress will depend on improving MOCVD control, epitaxial alignment, and data-driven optimization to produce large, defect-free films with high electronic performance.

### *Defects and oxidation in indium selenides*

As predicted by the In-Se-O phase diagram, indium selenides can undergo oxidation to finally form $In_2O_3$. When ultrathin InSe is exposed to pure oxygen or air at room temperature, *p*-type doping and current hysteresis behaviors can arise[84,85], making it difficult to achieve the desired high conductivity in transistor applications. Furthermore, the oxidative behavior of InSe depends on the degree of defects, such as Se vacancies,[86] and on the film thickness[60]. In the case of $In_2Se_3$, exposure to oxygen and moisture can induce the formation of an amorphous $In_2Se_{3-3x}O_{3x}$ surface layer, a process that is further accelerated under light illumination. Once this oxidation takes place, Se hemisphere particles can form on the oxidized $In_2Se_{3-3x}O_{3x}$ surface[87], which might act as trap center and undesired doping. Notably, the oxidation characteristics depend on the surface coordination environment: octahedral coordination of atoms in both α and β phases is less stable than tetrahedral coordination in α phase, thereby facilitating the formation of these Se particles[87]. The resultant $In_2Se_{3-3x}O_{3x}$ layer can self-passivate the surface by limiting oxidation to just a few nanometers in thickness, suggesting that thicker $In_2Se_3$ can exhibit reduced oxidation[87], thereby preserving the underlying crystal quality and carrier transport properties.

Consequently, optimizing the surface chemistry of indium selenides and developing protective coatings that preserve their electronic properties are imperative. The most widely used method to retard the oxidation of indium selenides involves fabricating encapsulation layers like $HfO_2$ by atomic layer deposition, which effectively isolates the crystals from moisture and oxygen in the environment and improves its stability[9]. One can also conduct dry oxidation to form a nonstoichiometric, self-limiting $InSe_{1-x}O_x$ oxide layer[60]. The formation of dry oxide layer retards further oxidation of the underlying InSe, thereby enabling a high two-probe field-effect mobility of greater than 450 $cm^2V^{-1}s^{-1}$ at 300 K for 13-nm-thick InSe in its field-effect transistor (FET)[60]. For device fabrication, it is essential to use an air-stable fabrication method compatible with photo-beam or electron-beam lithography. In those processes, baking the e-beam resists at 110 °C instead of the usual 180 °C or higher, and then quickly placing it in a high-vacuum environment, helps to minimize the formation of free radicals that degrade InSe contact interface[9].

In addition to the native oxides, intrinsic defects such as Se vacancies in indium selenide structures mediate electronic doping characteristics (**Supplementary Table 4**). Se vacancies generally act as deep-level traps or shallow donors, contributing to *n*-type behavior in In-rich conditions. Conversely, In vacancies serve as shallow acceptors under Se-rich environments and can induce *p*-type conduction in α-$In_2Se_3$ or Se-rich InSe[88,89]. Several strategies have been experimentally explored to engineer desired transport characteristics using extrinsic doping (**Supplementary Table 4**). For example, Sn or I substitution introduces *n*-type doping with improved conductivity[90,91], while Y substitution can induce a semi-metallic state in InSe[9]. These atomic heterogeneities can strongly influence the carrier concentration, recombination dynamics, and transport mobility, which are key metrics for logic devices such as future complementary FETs, which are the transistors that enable low-power and high-speed digital circuits by integrating both *n*- and *p*-type channels. Overall, understanding and controlling oxidation and defect chemistry are crucial for realizing stable, high-mobility indium selenide devices suited for electronic and memory applications.

## InSe Low-Power Transistors

Transistors are fundamental and critical components in modern electronics, driving the research community and semiconductor industries since their invention in the mid-20th century. 2D semiconductors offer an exciting path toward transistor scaling beyond the conventional 5 nm node. Their atomic-scale thickness of a few nm greatly improves electrostatic control, which makes them promising for physical sub-10 nm gate lengths. 2D channels have dangling-bond-free surfaces that reduce interface traps and variability, and they can be integrated in 3D device architecture, such as vertically stacked nanosheet or nanowire FETs for enhanced drive current and higher device density. Even with its simple planar structure, a metal-oxide semiconductor FET (MOSFET) using a 2D semiconductor channel demonstrates the potential to eventually replace conventional Si channels[4,92,93](**Fig. 3a**). Nevertheless, major hurdles remain, including finding the most viable channel material options among the versatile 2D semiconductors. InSe stands out as a promising material for logic device fabrication due to its high carrier mobility, low thermal conductivity, and ballistic transport.

Beyond conventional thermionic MOSFET operation, 2D semiconductors also enable novel steep-slope device concepts that can reduce the switching power below the limit of traditional FETs by overcoming the ~60 mV $dec^{-1}$ subthreshold slope (SS) limit at room temperature. Their exceptionally high carrier mobility makes 2D semiconductors particularly appealing for these emerging steep-slope transistors, while their van der Waals surfaces and easy integration offer extensive opportunities for band engineering and device design. For example, in TFETs, 2D heterojunctions with type-III band alignment can promote band-to-band carrier tunneling and facilitate sub-thermionic (<~60 mV $dec^{-1}$) switching, albeit with ongoing challenges in boosting the on-state current (**Fig. 3b**). These breakthroughs suggest that high-mobility 2D semiconductors, when combined with advanced device physics approaches like tunneling and 3D integration, could be key to continuing transistor scaling and enabling ultra-low-power operation in next-generation electronics.

## *FETs with 2D InSe nanosheet channels*

The most attractive feature of InSe for electronic applications is the high electron mobility. Even early-stage, non-optimized InSe transistors exhibit mobilities higher than what is achievable with single-crystal TMDs[94-96]. InSe transistor performance is improved by employing smooth polymer substrates[7,97,98], passivation by controlled oxidation[60,99,100], indium capping[8], and encapsulation with hBN[6,101,102]. Room temperature field effect mobilities in these devices can routinely exceed 1,000 $cm^2V^{-1}s^{-1}$; competing with state-of-the-art silicon transistors[6-8].

This high mobility of InSe is attributed to an exceptionally high thermal velocity exceeding $2\times10^7$ cm $s^{-1}$ (ref.[9]), which surpasses that of all semiconducting TMDs and silicon at equivalent thicknesses[9] (**Fig. 1b**). This high thermal velocity enables electrons to travel through the transistor channel with minimal scattering when the channel length is short enough and ohmic contact exists. As a result, carriers traverse the channel without undergoing substantial scattering with phonons or impurities, preserving their energy and momentum, which is critical for reducing power dissipation and ensuring faster operation (**Fig. 3a**, right). This is important because, at ultra-short channel lengths, 2D material FETs are typically limited by the thermal velocity of the channel material, which is determined solely by the material's effective electron mass[10,32,33]. The superior thermal velocity of InSe owes to its small effective mass of $\sim 0.1m_0$ down to monolayer thicknesses. However, the small effective mass can trigger the disadvantage of increasing the probability of source-drain tunneling in the off state, known as the leakage current[103]. Therefore, to realize InSe-based FET devices with a high on-to-off current ratio, it is vital to suppress the leakage current. One strategy is to select the appropriate channel InSe thickness and thereby increase the effective mass and bandgap due to quantum confinement effects, which reduces the probability of tunneling. Additionally, integrating high-$k$ gate dielectrics or adopting gate-all-around geometries enhances gate control, thereby minimizing tunnelling leakage current through improved electrostatic control of the channel.

By employing contact engineering, encapsulation with high-$k$ dielectric, and scaled channel lengths below 20 nm, ballistic transistors utilizing 2.4 nm-thick InSe have been demonstrated[9]. The device, with a 10 nm channel length, achieves ohmic contacts through Y-doping-induced phase transitions in InSe. By minimizing phonon scattering and leveraging its high thermal velocity, the InSe ballistic transistors demonstrate exceptional transfer characteristics, including a SS of ~75 mV $dec^{-1}$ (**Fig. 3a**, **3c**, and **3d**) and a record-high transconductance of 6 mS (ref.[9]) (**Fig. 3e**), better than other 2D material MOSFETs.

Still, most InSe FETs typically exhibit intrinsic *n*-type behavior, primarily attributed to native point defects, such as In interstitials and Se vacancies (**Supplementary Table 4**), as well as Fermi-level pinning at the metal-semiconductor interface (**Supplementary Fig. 1** and **Supplementary Note 1**). To implement the complementary FET architecture, which necessitates possessing both *n*- and *p*-type channels, the fabrication of high-performance *p*-type 2D InSe FET is necessary. One effective strategy involves using electrical contact of high-work-function metal, Pt, configured in a van der Waals interface, which can mitigate Fermi-level pinning (**Supplementary Fig. 1**) and enhance hole injection in FETs[104].

***Tunnel FETs based on InSe/Si heterojunction***

TFET composed of 2D InSe and 3D Si (**Fig. 3b**) exhibits sub-60 mV $dec^{-1}$ operation, an on-state current density of ~0.3 μA $μm^{-1}$, and an on/off current ratio of ~$10^6$ (ref.[20]). These remarkable results are attributed to the type-III band alignment formed between InSe and Si, which establishes an optimal offset in their conduction and valence bands. Moreover, the atomic-scale thickness of InSe not only maximizes gate electrostatic control but also, owing to its van der Waals surface, provides the clean and thin channel essential for TFETs. Consequently, when suitable gate and drain voltages are applied, band-to-band tunneling is readily induced.

The subthreshold swing of InSe TFETs[20] is comparable to that of BP-based TFETs[107] and superior to that of 3D InAs-based TFETs[108] (**Fig. 3d**). However, the current density of InSe TFET, which is less than 1 mA $μm^{-1}$ (ref.[20]), is still lower than those of Si- or 2D semiconductor-based MOSFETs[9,109,110] (**Fig 3d, e**). These characteristics suggest that there is substantial room for improvement in achieving high on-state conductance. Thus, it is necessary to find better material options for semiconductor with bandgap compatible with InSe, and to implement automated, large-area processes that can optimize characteristics of interfaces at junctions, contacts, and dielectrics.

***Flat bands of InSe and their detection in FETs***

In addition, InSe FETs provide an excellent platform for studying new physical properties of 2D InSe. One notable discovery involves detecting InSe's flat band through electrical measurements[105] (**Fig. 3f**). The measurement of the flat band can be conducted using a FET with a simple but effective design: it is sandwiched InSe between two hBN insulating layers and uses graphene for the source and drain contacts (**Fig. 3f**, left). By adjusting the gate voltage, one can move InSe's Fermi level from the conduction band, through the bandgap, and into the valence band region. Hence, out-of-plane tunneling current using carriers generated by visible laser light can be measured. When the Fermi level reaches the van Hove singularity near the valence band (**Fig. 3f**, right), the density of states increases sharply, causing a distinct spike in tunneling current. This effect becomes particularly noticeable as the tunneling mechanism shifts from direct to Fowler-Nordheim tunneling at the flat band. Compared to traditional measurement techniques such as angle-resolved photoemission spectroscopy, scanning tunneling microscopy, and scanning tunneling spectroscopy, this electrical detection method is remarkably simple[105]. Because flat bands typically exhibit large effective masses and strong electron-electron interactions, InSe might display interesting phenomena like magnetic and superconducting phase transitions. Additionally, the transistor structure enables studying spin polarization and chirality effects in 2D InSe by measuring photo-excited holes tunneling through the hBN barrier under magnetic fields[125].

***Ballistic avalanche InSe-BP FETs***

By leveraging InSe's ballistic transport characteristics and van der Waals properties, one can form a high-performance FET based on heterostructures composed of InSe and other 2D or 3D semiconductors (**Fig. 3g**). Depending on how carriers transport across these semiconductors, such heterostructures can be utilized in devices such as impact ionization transistors or tunneling transistors.

For instance, by using InSe as an *n*-type semiconductor and BP as a *p*-type semiconductor, both electrons and holes can be sufficiently accelerated within the heterostructure to drive impact ionization[106] (**Fig. 3g**). As a result, unlike the conventional impact ionization-driven avalanche phenomenon observed in bulk semiconductors, a ballistic avalanche phenomenon can be observed in vertical InSe/BP heterostructures. While typical avalanche devices require high voltages in the order of tens of volts or a long impact ionization region, the InSe/BP heterostructure enables avalanche breakdown at voltages below ~1 V.

Moreover, the devices exhibit excellent noise performance with a low avalanche threshold below 1 V, and an ultra-steep SS of 0.25 mV $dec^{-1}$, demonstrating their potential for high-sensitivity avalanche photodetectors and transistors[106].

## Non-Volatile Memory Device Applications

Ferroelectricity in 2D materials has attracted much research interest due to its potential applications in energy-efficient computing platforms when applying the non-volatile properties of the ferroelectric polarizations[111]. When coupled with the van der Waals nature of 2D materials, in particular, ferroelectricity offers an innovative approach to address the scaling limitations and performance degradation at reduced thicknesses faced by traditional 3D ferroelectrics[14,111,112]. Moreover, van der Waals layered InSe and $In_2Se_3$, even with their atomic-scale thickness and a bandgap of > 1.41 eV, exhibit either sliding ferroelectricity or spontaneous ferroelectricity despite not being insulators. Consequently, the existence of ferroelectricity in indium selenides enables the integration of strong polarization responses with functionalities derived from their semiconducting properties, facilitating the development of compact, energy-efficient devices for in-memory, in-sensory computing, neuromorphic systems, and optoelectronics[14,111,112]. Furthermore, the high Curie temperature of $In_2Se_3$ makes it especially attractive, as it allows for maintaining stable operation under high temperatures generated by Joule heating. Therefore, InSe and $In_2Se_3$ can be applied to ferroelectric semiconductor junctions (FSJs)[19,50,113] and FeSFETs[12,18,114] for future ultra-high-density, low-power memory applications.

### ***$In_2Se_3$ ferroelectric semiconductor junctions***

Ferroelectric tunnel junctions (FTJs) are next-generation memory devices that utilize electron tunneling through an ultrathin ferroelectric insulating film, exhibiting nonvolatile resistance changes depending on the polarization orientation. A typical FTJ adopts a metal–ferroelectric–metal configuration, in which a nanometer-thick ferroelectric layer serves as the tunneling barrier and is sandwiched between two metal electrodes, and an oxide or semiconducting interlayer is sometimes inserted between metal and ferroelectric. In this structure, electrons tunnel quantum-mechanically through the ferroelectric barrier, and the reversal of ferroelectric polarization shifts the barrier height and width, thereby altering the tunneling current in a nonvolatile manner. This phenomenon, referred to as tunneling electroresistance, is quantified by its on/off ratio, which is a key figure of merit for FTJ memory operation.

For ferroelectric semiconductor $In_2Se_3$, rather than an ferroelectric insulator, crossbar two-terminal FSJ devices share the same metal–ferroelectric–metal structure of FTJs (**Fig. 4a**). However, polarization-dependent charge transport can occur via thermionic emission over a Schottky barrier, rather than tunneling electroresistance in FTJs (**Fig. 4b**). This key distinction enables FSJs to achieve high on/off ratios over ~$10^4$, improved thermal stability, and reduced leakage without requiring ultrathin films in various $In_2Se_3$ FSJs[19,113,115]. Furthermore, due to the semiconducting nature of $In_2Se_3$, partial polarization switching confined near the contact interface can occur under localized electric fields[116]. Compared with typical insulators, the smaller bandgap of $In_2Se_3$ facilitates non-volatile modulation of the Schottky barrier height without full switching across the entire film thickness, offering an additional degree of partial ferroelectric control in FSJ. Therefore, FSJs can exhibit potential for multi-state applications in analog synaptic devices.

Contributed to van der Waals stacking of different materials, the FSJ structures of metal/interlayer/α-$In_2Se_3$/p+Si[ref.19], metal/h-BN/α-$In_2Se_3$/graphene[113], metal/α-$In_2Se_3$/$MoS_2$[ref.115] have been proposed. The insertion of an interlayer can provide asymmetry of the band alignment, while modulating effective Schottky barrier height[19,113,115] and suppressing the thermionic current leakage[115]. For example, the metal/α-$In_2Se_3$/$MoS_2$[ref.115] device exhibits both room-temperature negative differential resistance effect and high electroresistance exceeding $10^4$ simultaneously, providing temperature-independent transport. Ferroelectric van der Waals metal, $WTe_2$, has been used as metal electrode for $In_2Se_3$ FSJs in $WTe_2$/α-$In_2Se_3$/metal[117]. The use of $WTe_2$ allows for the high on/off ratio of $10^5$, with a switching voltage of less than 2 V. In addition, multiple resistance levels have been observed due to the pinning effect of the $WTe_2$/α-$In_2Se_3$ interface on the upward polarization of α-$In_2Se_3$. This pinning effect permits the polarization of α-$In_2Se_3$ to maintain a partially switched polarization state, resulting in an intermediate resistance level.

Despite this potential, however, an on/off ratio of $In_2Se_3$ FSJ[19,50,113,115] is limited to ~$10^5$, which is relatively low compared to other FTJs. The state-of-the-art bulk ferroelectric-based FTJ composed of oxide heterostructures, exhibits an on/off ratio of ~$10^8$ (ref.[118]), and the van der Waals ferroelectric $CuInP_2S_6$-based FTJ achieves ~$10^{10}$ at RT (ref.[119]), both higher than that of $In_2Se_3$ FSJ[19,50,113,115] (**Table 1**). More importantly, there remain significant challenges for the large-area integration of $In_2Se_3$ FSJs into crossbar array devices within Si back-end-of-line compatible fabrication processes and large-scale integration technologies.

### *InSe and $In_2Se_3$ ferroelectric semiconductor FETs*

Compared with two-terminal FTJs or FSJs, three-terminal ferroelectric FETs (FeFETs) offer multiple advantages in terms of operating mechanism and practical utility[120]. For instance, two-terminal ferroelectric devices typically ascertain the memory state by applying voltage and measuring the resulting current, which can lead to destructive readout if the polarization is altered or partially lost during measurement[121]. In contrast, in a three-terminal FeFET architecture, one can measure the channel doping state indirectly by simply adjusting the gate voltage, thus enabling non-destructive readout[120,122]. Because the read and write pathways are physically separated, these transistor-based devices provide higher device stability and better data retention. Additionally, the channel conductivity is modulated through the gate, while ferroelectric polarization maintains non-volatility, making it more straightforward to combine logic and memory functionality within a single device[123]. Thus, FeFETs are well suited for next-generation non-volatile memories, neuromorphic computing, and logic-in-memory applications, thanks to their fast-switching characteristics, potential for high-density integration, and low power consumption. By finely tuning the gate voltage, these devices can realize multi-level channel current control, synaptic weight updates, and independent readout signals.

In the case of InSe and $In_2Se_3$, both materials can exhibit ferroelectric behavior yet have a relatively small bandgap of > 1.41 eV compared with bulk ferroelectric insulators (**Fig. 1d**). This property has attracted attention for potential applications as FeSFETs[12,18,124,125] (**Fig. 4c**). Unlike conventional FeFETs that use a ferroelectric layer as the gate dielectric, InSe and $In_2Se_3$ allow the use of a standard gate dielectric while still enabling polarization control in the channel. Because the channel itself is ferroelectric, reorienting the polarization critically alters the channel doping concentration and conductivity (**Fig. 4d**). Depending on the effective oxide thickness, different portions of the channel's top or bottom surfaces can be polarized, and the electric field can penetrate to varying depths[18]. Consequently, the transfer characteristics can exhibit clockwise or counterclockwise hysteresis, forming a memory window, and because this polarization is preserved even when power is removed, these transistors can serve as non-volatile memory that integrates logic and memory in a single device.

In conventional FeFETs employing a bulk ferroelectric gate, incomplete screening of polarization along the channel can degrade memory properties[126]. In contrast, ferroelectric-semiconductor channels, such as InSe and $In_2Se_3$, can self-screen the polarization charges with mobile carriers, thereby reducing interface trapping and gate leakage and improving retention[18,111]. This advantage can be explained in terms of the depolarization field. The depolarization field is an internal electric field opposite to spontaneous polarization that arises when bound charges at the ferroelectric interfaces are only partially compensated[127]. For a typical FeFET with a metal-ferroelectric-interlayer-semiconductor gate structure, the depolarization field can be severe when a thin, low-*k* interfacial oxide like $SiO_2$ in a TiN/HZO/$SiO_2$/Si stack separates the ferroelectric from the channel; as the magnitude of the depolarization field approaches the coercive field, retention degrades rapidly because the polarization can no longer remain stable against this self-imposed field. FeSFETs based on atomically thin ferroelectric semiconductors can avoid this limitation because the channel itself is ferroelectric. The van der Waals interface supplies nearly ideal charge compensation without dangling bonds, and the FeSFET device stack typically lacks a distinct interfacial oxide, eliminating the voltage drop that would otherwise generate a strong depolarization field. Retention can be further enhanced by combining this intrinsic advantage with extrinsic engineering, such as adopting an ultrathin, high-*k* gate dielectric to shorten the screening length, and by applying modest electrostatic or chemical doping to the ferroelectric channel, ensuring a sufficient free-carrier density for rapid charge compensation. In short, 2D FeSFETs hold the promise of mitigating the depolarization-field-limited retention that has long hindered bulk-oxide FeFETs.

In addition, switching speeds can reach ~40 ns in $In_2Se_3$-based FeSFETs[128]. This fast-switching behavior could possibly arise from the intrinsic coupling between ferroelectricity and semiconducting nature of α-$In_2Se_3$: unlike traditional FeFETs where only polarization-bound charges participate, the ferroelectric semiconductor channel supports both bound and mobile charges[18,111,128]. The presence of mobile carriers enables the formation of internal electric fields that assist and stabilize the polarization switching, thereby enhancing switching dynamics. In this configuration, the internal field compensates for depolarization and reinforces domain alignment even under moderate gate voltages.

However, the switching timescale is still slower than that of non-2D ferroelectrics operating in the sub-nanosecond regime, such as Pb(Zr,Ti)$O_3$[ref.129]. Therefore, realizing further acceleration of switching speed in 2D FeSFETs remains an open challenge. This limitation could be particularly attributed to gate capacitance limitations and incomplete polarization caused by relatively thick dielectrics or weak vertical electric fields. As such, continued optimization of gate stack engineering—such as reducing the effective oxide thickness to enhance the vertical electrical field—will be essential for pushing the switching speed.

Furthermore, device-level validation of interface stability, operating temperature ranges, and long-term endurance is also becoming increasingly important, given that current $In_2Se_3$ FeSFETs[12,18,124] often show endurance ~$10^4$ cycles and retention on the order of $10^5$ seconds (**Table 1**). To ensure robust performance, it is essential to optimize selection of metal contacts and gate dielectrics to minimize contact resistance, trap states, and device degradation. In this regard, employing 2D ferroelectric metal contacts similar to work with $WTe_2$/α-$In_2Se_3$/metal crossbar devices[117] combined with Fermi-level-depinned metal contacts (**Supplementary Note 1**) will enable more stable, low-resistance contact interfaces and enhanced endurance in future FeSFET.

In addition, the demonstration of endurance over $10^{11}$ cycles in FeFETs based on the sliding ferroelectricity of bilayer BN (ref.[130]) indicates that sliding ferroelectricity in InSe can offer enhanced robustness against fatigue. Here, polarization reversal proceeds through an in-plane interlayer shear, described as a shear-mediated phase transformation[131], which transforms pre-existing, movable ferroelectric domain boundaries rather than repeatedly nucleating and annihilating out-of-plane domains. Because this shear occurs across weak van der Waals gaps, it largely suppresses the creation of dangling bonds, oxygen vacancies, and other interface defects that usually accumulate at the junction between ferroelectric and semiconductor, resulting in reliability issues and narrowing the memory window in conventional FeFETs[132]. Therefore, sliding ferroelectrics maintain high endurance through an in-plane sliding mechanism[130]. Though this intrinsic “defect-tolerant” switching pathway explains the high cycling endurance observed thus far, systematic studies are still needed to correlate native defects, possible interlayer slip pinning, and long-term retention, thereby validating the ultimate reliability limits of InSe-based sliding ferroelectric memories.

**Ferroelectric devices for in-memory computing**

Indium selenide-based ferroelectric memory devices can provide new opportunities for beyond-von Neumann computing architectures. These devices enable simultaneous storage and processing of information within the same physical element, thereby overcoming the memory-logic bottleneck in conventional Si-based semiconductor systems. The non-destructive readout and low switching energy further enhance their applicability in low-power embedded systems. Furthermore, $In_2Se_3$-based FeFETs can perform Boolean operations by utilizing polarization-dependent conductance states, thereby enabling logic-in-memory or compute-in-memory architectures. For example, ferroelectric polarization can modulate channel resistance to represent different logic states, allowing the realization of AND, OR, and majority logic gates directly within memory arrays. Such architectures benefit from fine-grained parallelism and reduced interconnect overhead, which are crucial for data-centric workloads such as matrix-vector multiplication and real-time signal processing.

Beyond digital logic, indium selenide memory devices have shown promising characteristics for neuromorphic computing. The analog tunability of ferroelectric polarization in $In_2Se_3$ FeFETs enables gradual conductance modulation, mimicking biological synaptic plasticity[133-135]. Spike-dependent programming and long-term potentiation and depression behaviors make suitable artificial synapses in hardware neural networks[133-135]. Additionally, the comparatively wider bandgap and optical activity of $In_2Se_3$, especially with respect to Si, allow integration with photodetectors to realize retinomorphic systems that process visual stimuli at the sensor level with minimal energy overhead.

## Outlook

The development of van der Waals indium selenides for advanced logic and memory applications remains at an exciting early stage compared to established silicon and compound semiconductor technologies. The key engineering hurdle is to tame the material's complexity while scaling from record-setting material properties and device performance metrics in exfoliated flakes to wafer-level manufacturing. MOCVD can ultimately deliver high-purity InSe and $In_2Se_3$ films on > 200 mm wafers. Achieving this target will require the strict suppression of oxide impurities, vacancies, and other defect-forming mechanisms; robust oxidation-resistant encapsulation that survives backend temperatures below 450 °C; and contact electrodes that decisively eliminate Fermi-level pinning, allowing both *n*-type and *p*-type operation to become feasible.

Once these large-area process challenges are met, architectural attention will shift to exploiting the interesting characteristics of indium selenides. A compelling concept is a monolithically stacked logic-in-memory block[136], in which nanometer-scale gate-all-around InSe transistors on the lower tier provide fast logic. In contrast, an upper-tier array of ferroelectric semiconductor transistors or junctions supplies non-volatile storage and analog multiply-accumulate functionality. Demonstrating such heterogeneous integration at practical densities would show that indium-selenide electronics can coexist with, or augment, conventional Si for system-on-chip platforms.

Still, the most immediate research priority should be improving the endurance of $In_2Se_3$ ferroelectric devices beyond their current $<10^4$ cycle limitation toward the goal of $>10^{11}$ cycles, the threshold necessary for reliable non-volatile memory operation. To enhance the cycling endurance and reliability of $In_2Se_3$-based memories, efforts should focus on reducing intrinsic defect density, particularly Se vacancies and In interstitials, through controlled doping and defect passivation strategies such as oxygen incorporation. As the technology matures, vertically integrating dozens of indium-selenide—each only a few atoms thick and separated by self-aligned metallic interconnect planes—will become possible. This monolithic 3D integration promises memory densities while sitting directly atop logic, eliminating the von Neumann bottleneck and the area penalties of through-Si vias.

Looking further ahead, compositionally tuned 2D $In_xGa_ySe_z$ alloys could unlock reconfigurable band-gap engineering and other features that push the frontiers of adaptive electronic and neuromorphic systems[69-72]. Alloying InSe with GaSe to form $In_{1-x}Ga_xSe$ compounds would enable continuous tuning of band structures, effective masses, and optical properties beyond what is possible with phase and thickness control alone. These ternary alloys could offer optimized combinations of mobility, bandgap, and stability tailored for specific applications from high-performance logic to photodetection. Furthermore, the In-Ga-Se system potentially harbors entirely new phases with unique properties at specific compositional ratios, analogous to discoveries in other ternary chalcogenide systems. The development of compositional gradient structures or vertical heterostructures within this material family could yield novel quantum phenomena and device functionalities impossible in binary compounds.

The transformative potential of indium selenides ultimately depends on continued cross-disciplinary collaboration between materials scientists, device physicists, process engineers, and system architects. By addressing the full spectrum from fundamental properties to practical implementation challenges, this emerging material platform can fulfill its promise as a cornerstone of next-generation, low-power computing technologies.

## Acknowledgements

This work was supported by the National Research Foundation of Korea (NRF) grant funded by the Korea government (Ministry of Science and ICT, MSIT) (No. RS-2025-00516532). S.S.,

W.L. and H.S.S. acknowledge partial support for this work by the Institute for Basic Science (IBS-R036-D1). M.A. and N.G. acknowledge the funding from the Air Force Office of Scientific Research (AFOSR) under Award No. FA9550-24RYCOR011. D.J. and S.S. acknowledge primary support from the National Science Foundation, Future of Semiconductors (FuSe) program award number 2328743. D.J. and S.S. also acknowledge partial support from the Office of Naval Research (ONR) Nanoscale Computing and Devices program (N00014-24-1-2131) as well as the Air Force Office of Scientific Research (AFOSR) GHz-THz program grant number FA9550-23-1-0391

## Author contributions

D.J., S.S. and N.G. conceived the idea of the content. All authors researched data for this article. S.S. and M.A. contributed substantially to the discussion of the content and wrote the article, with input from W.L. and H.S.S. All authors reviewed and edited the content before submission.

## Competing interests

The authors declare no competing interests.

### Key Points

- Indium selenides uniquely combine exceptional electron mobility of >1,000 $cm^2V^{-1}s^{-1}$, high thermal velocity of >$2\times10^7$ cm $s^{-1}$, and diverse phase-dependent properties that make them superior to other 2D materials for ultra-scaled transistors and memory devices[6,9].
- The polymorphic nature of InSe (β, γ, ε phases) and $In_2Se_3$ (α, β, β' phases) enables precise tuning of electronic and ferroelectric properties, offering unprecedented design flexibility for device engineers compared to conventional semiconductors with fixed properties[11-13].
- Ballistic InSe transistors with 10 nm channels have achieved record-high transconductance of ~6 mS and current densities that outperform other 2D materials and approach theoretical limits, validating indium selenides' potential for post-silicon logic devices[6,9].
- The robust ferroelectricity in $In_2Se_3$ and emerging sliding ferroelectricity in InSe enable non-volatile memory functionality integrated with semiconducting properties—a combination unavailable in conventional materials that could fundamentally transform computing architectures[15,18,19].

- Addressing the challenges of scalable synthesis, phase control, and oxidation prevention represents the critical path toward commercial implementation of indium selenide-based technologies for next-generation, low-power computing beyond the silicon-based semiconductor chips.

**Table 1. Benchmarking of electrical performance metrics in 2D ferroelectric devices based on $In_2Se_3$.†**

| Material | Device type | Contact metals | Channel thickness | Switching speed | On/off ratio | Memory window (V/nm) | Retention time (sec) | Endurance (cycles) | Ref. |
|---|---|---|---|---|---|---|---|---|---|
| α-$In_2Se_3$ | Lateral gate-FeFET | Ti/Au | 68 nm | N/A | $10^4$ | 0.39 | $3\times10^4$ | $10^5$ | 137 |
| | FeSFET | Cr/Au | Few-layer | N/A | $10^3$ | N/A | $10^4$ | $10^4$ | 12 |
| | | - | 40 nm | 40 ns | $10^5$ | 0.15 | $10^3$ | 500 | 128 |
| | | Ti/Pt | 47 nm | 1 s | $10^3$ | 0.27 | $6\times10^4$ | 1200 | 124 |
| | | Ti/Au | 20-50 nm | 100 ns | $10^5$ | 0.4 | 150 | 250 | 48 |
| | | Ni | 30 nm | N/A | $10^6$ | N/A | 0.1 | $10^4$ | 138 |
| | FSJ | $MoS_2$ or Ti | 12 nm | N/A | $10^4$ | N/A | $5\times10^3$ | $4\times10^3$ | 115 |
| | | Au or Ti | 40 nm | 10 ns | $10^3$ | N/A | $5\times10^3$ | 100 | 139 |
| | | Au or $WTe_2$ | 26 nm | N/A | $10^5$ | 0.17 | N/A | N/A | 117 |
| $CuInP_2S_6$ | FTJ | Au/Cr | 4 nm | 10-50 μs | $10^7$ | 0.14 | $10^7$ | $5\times10^3$ | 140 |
| $MoS_2$ with Sc-doped AlN gate | FeFET | Ti/Au | Few-layer | N/A | $10^6$ | 0.3 | $10^5$ | $10^4$ | 141 |

†Abbreviations used: FeFET, ferroelectric field-effect transistor; FeSFET, ferroelectric semiconductor field-effect transistor; FSJ, ferroelectric Schottky junction; FTJ, ferroelectric tunnel junction; N/A, not applicable.

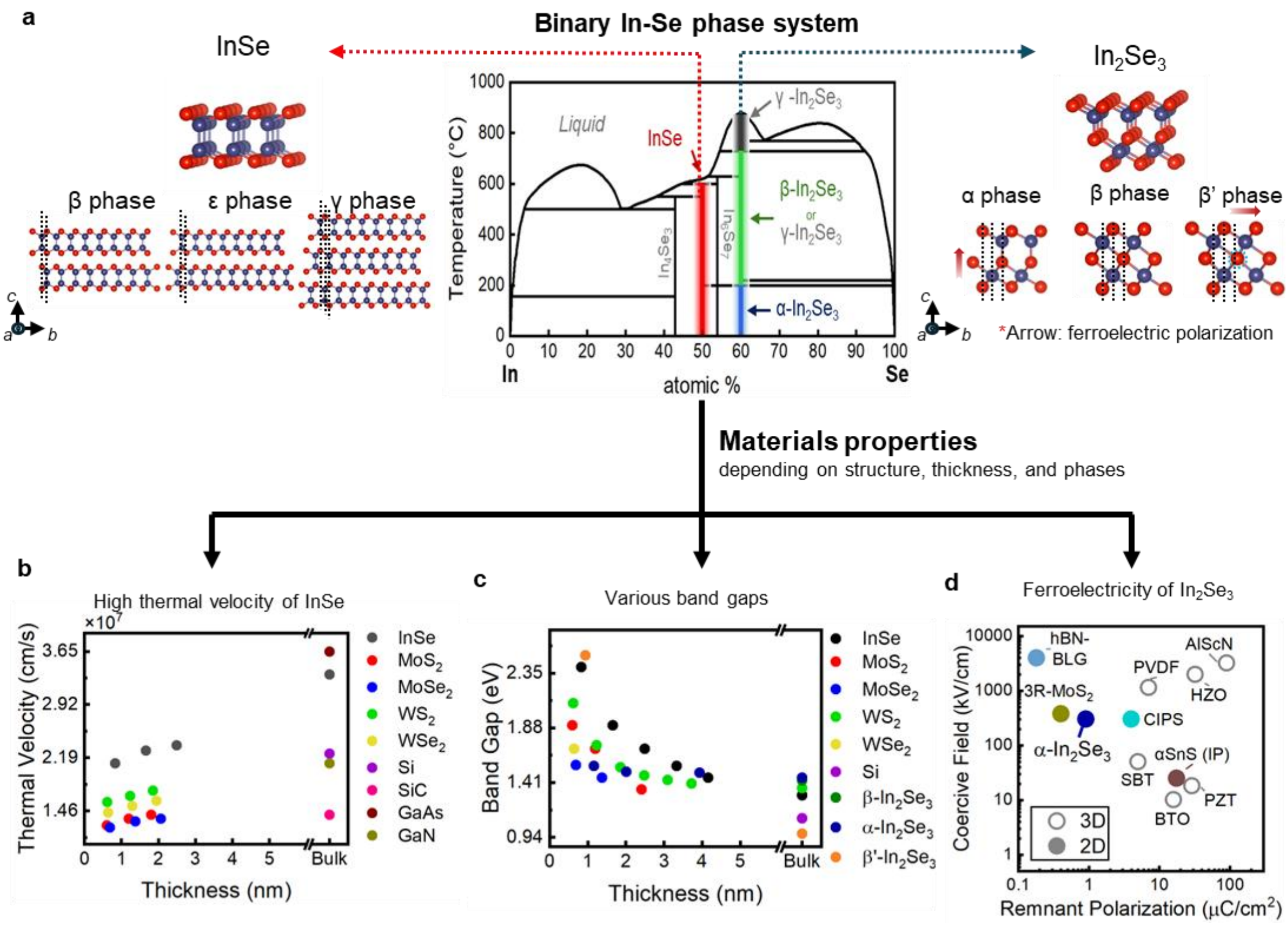

**Figure 1. Indium selenide structural phases and electronic properties.** (**a**) Schematic structures of InSe (left) and $In_2Se_3$ (right) showing their phase variants and stacking orders, with the binary In-Se phase diagram shown in the center. (**b**) Thermal velocity as a function of thickness for InSe compared with other semiconducting materials such as 2D transition metal dichalcogenides (TMDs), 3D compounds semiconductors, and 3D Si[9]. (**c**) Thick-dependent band gaps of the indium selenides compared with the semiconducting 2D TMDs and bulk Si[6,25,28,41,44]. (**d**) Coercive field and remnant polarization of α-$In_2Se_3$ compared with other 2D van der Waals ferroelectrics[131,142-147], such as bilayer graphene (BLG), α-phase SnS (αSnS), $CuInP_2S_6$ (CIPS), and rhombohedral-stacked $MoS_2$ (3R-$MoS_2$) (filled symbols) as well as conventional 3D ferroelectrics[148-151] such as Sc-doped AlN (AlScN), Poly(vinyldene fluoride) (PVDF), $Hf_xZr_{1-x}O_2$ (HZO), $SrBi_2Ta_2O_9$ (SBT), $BaTiO_3$ (BTO), and $Pb(Zr,Ti)O_3$ (PZT) (opened symbols). Here, all materials exhibit out-of-plane ferroelectricity, except for α-phase SnS, which has in-plane (IP) ferroelectricity.

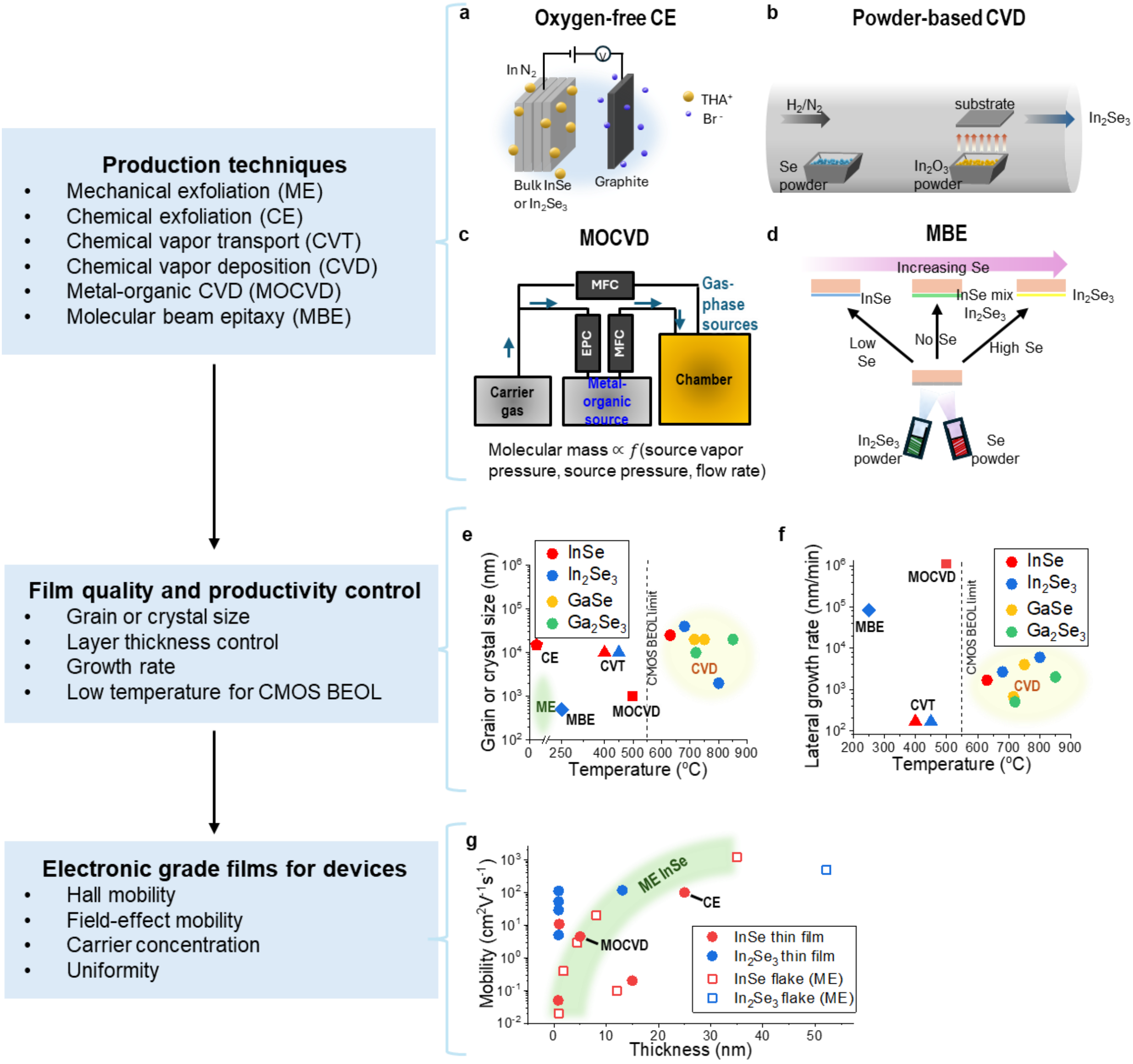

**Figure 2. Growth techniques for van der Waals semiconductors in the In-Ga-Se system.** (**a-d**) Representative synthesis methods for indium selenide thin films: (**a**) chemical exfoliation (CE), (**b**) powder-based chemical vapor deposition (CVD), (**c**) metal-organic CVD (MOCVD), and (**d**) molecular beam epitaxy (MBE)[80]. (**e,f**) Growth temperature dependence of (**e**) grain or exfoliated crystal size, and (**f**) lateral growth rate, summarized from reports on mechanical exfoliation (ME), CE, CVD[74,76,152-156], chemical vapor transport (CVT)[75], MBE[80], and MOCVD[11]. ME and CVD data are shown with green and yellow shading. (**g**) Thickness-dependent field-effect mobility of InSe and $In_2Se_3$ thin films mainly grown by CVD; MOCVD and CE are labeled, and ME samples appear as open squares with InSe highlighted in green[11,74,76,152-156].

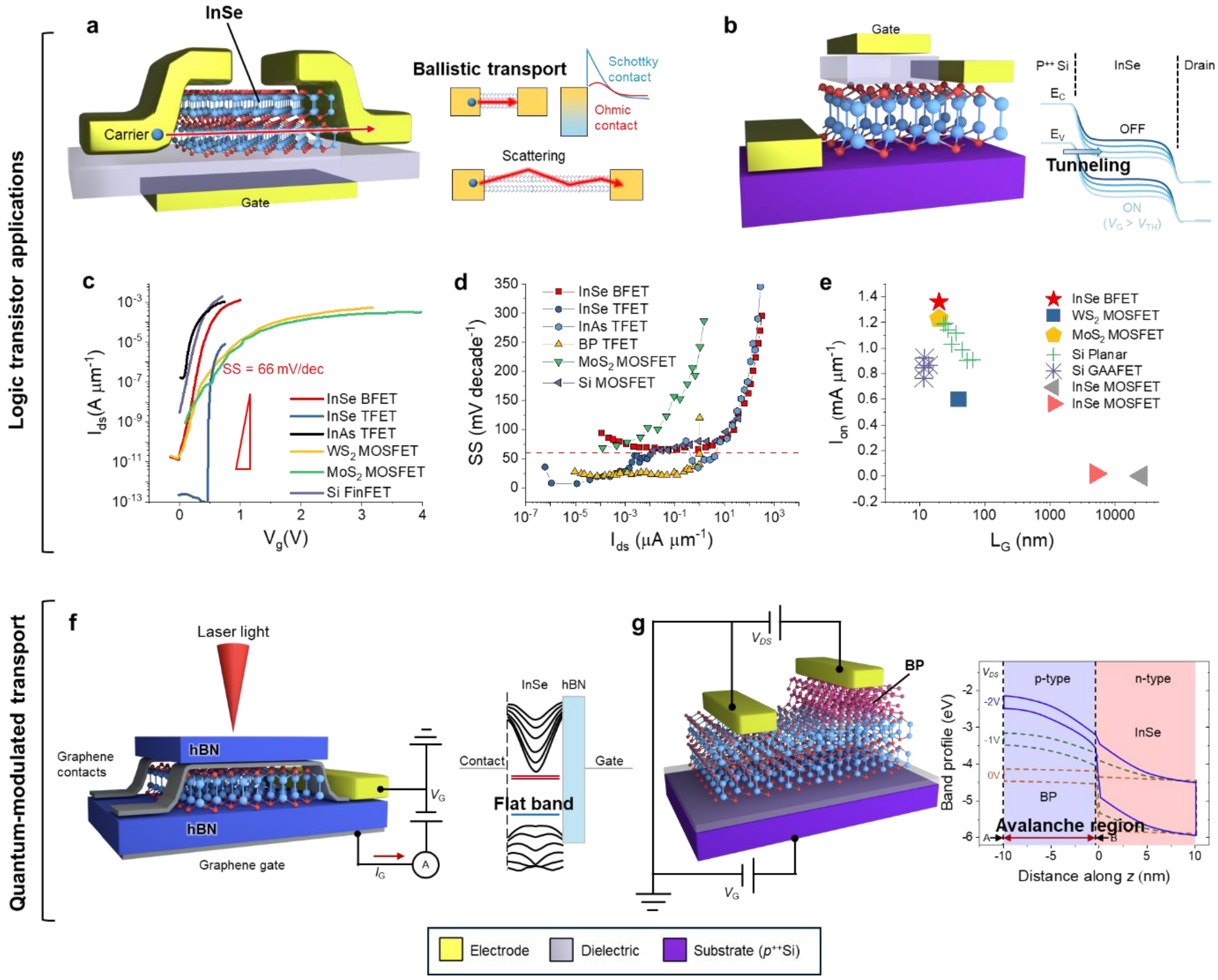

**Figure 3. InSe-based transistors and their transport mechanisms.** (**a**) Ballistic field-effect transistor (FET) with on InSe channel, where contact engineering and short channel enable ballistic rather than scattering-limited transport[9]. (**b**) Tunneling FET (TFET) employing InSe and $p^{++}$Si for band-to-band tunneling[20]. (**c-e**) Benchmarking plots of (**c**) transfer curves, (**d**) subthreshold swing (SS), and (**e**) on-state current density ($I_{on}$) versus channel length ($L_G$) for InSe-based FETs—ballistic FET (BFET), TFET, metal-oxide-semiconductor FET (MOSFET) and fin-shaped FET (FinFET)—compared with Si and other conventional semiconductor materials[9,20,107-110]. (**f**) Flat band observation through gate-source tunneling current ($I_G$) under laser illumination[105]. (**g**) Ballistic avalanche FET showing impact-ionization-driven carrier multiplication in InSe heterostructure[106].

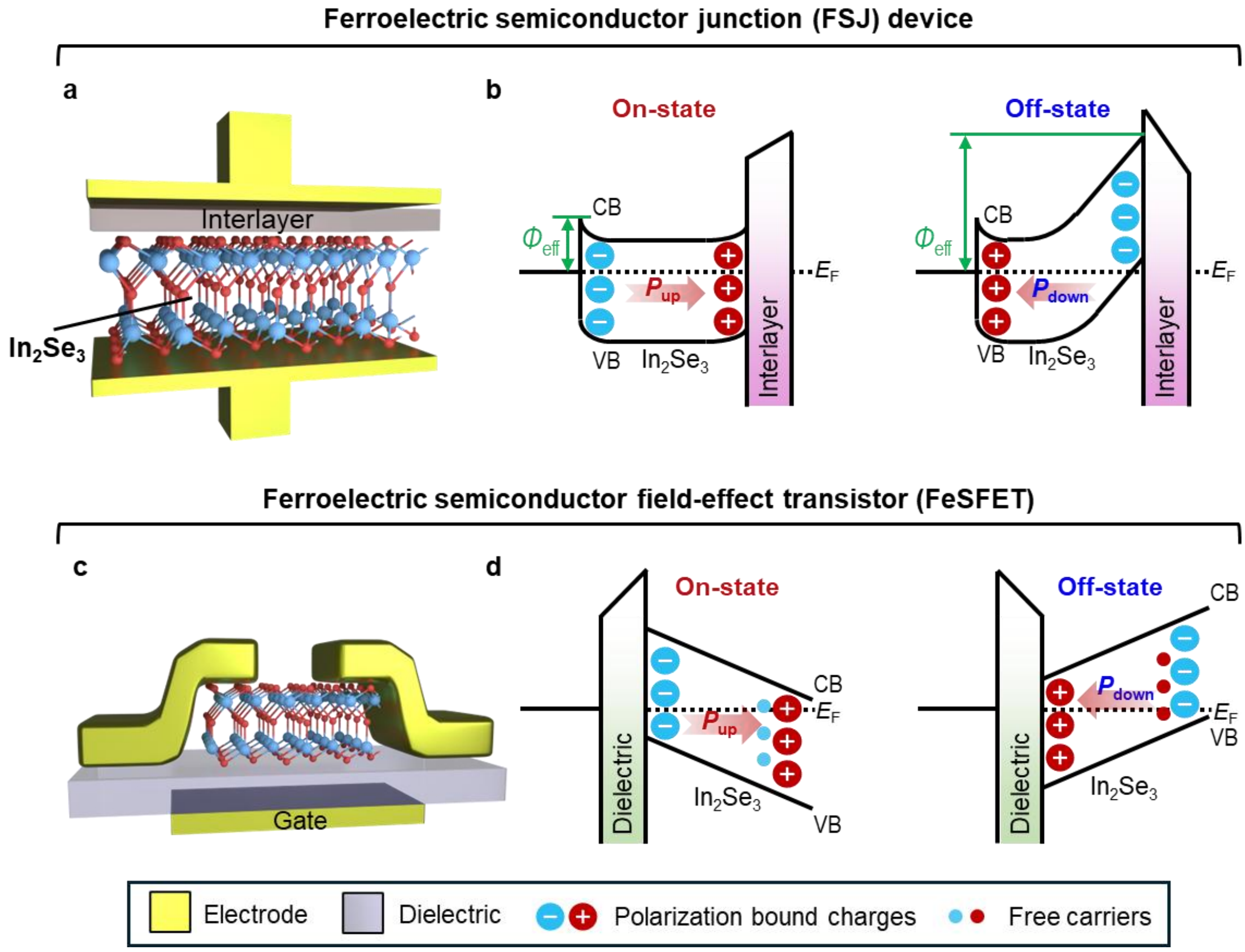

**Figure 4. Non-volatile ferroelectric memory devices based on $In_2Se_3$** (**a**,**b**) $In_2Se_3$-based ferroelectric semiconductor junction (FSJ) device with an interlayer. Illustrations of (**a**) the representative device structure, and (**b**) band diagram of the FSJ with each effective Schottky barrier height ($\Phi_{eff}$) at the on-state and off-state. (**c**,**d**) Ferroelectric semiconductor field-effect transistor (FeSFET) using $In_2Se_3$. (**c**) Device structure, and (**d**) Band diagram of a FeSFET in polarization up ($P_{up}$) and polarization down ($P_{down}$) states. $E_F$, Fermi level; CB, conduction band; VB, valence band.

**Box 1. Sliding ferroelectricity in 2D van der Waals materials**

Sliding ferroelectricity refers to a unique form of ferroelectricity that arises in layered 2D materials through the relative in-plane displacement, or sliding, between adjacent atomic layers. Unlike conventional ferroelectrics, where polarization switching typically results from ion displacement within a unit cell in response to an electrical field (panel **a**), sliding ferroelectrics exhibit polarization changes due to shifts in stacking order. This phenomenon is enabled by the weak van der Waals interaction between layers, which allows for reversible transitions between energetically distinct stacking configurations (panel **b**).

When layers in van der Waals 2D InSe slide relative to one another, the overall symmetry of the system breaks, leading to spontaneous out-of-plane polarization (panel **b**). Because multiple metastable stacking configurations can exist, these systems often exhibit multilevel polarization states *P*+ and *P*-, which manifest as staircase-like or asymmetric *P*-*E* hysteresis loops (indicated by red arrows in figure, panel **c**). This stepwise behavior also reflects the spatial inhomogeneity of the sliding process, where different domains within the crystal undergo stacking transitions at different threshold fields due to local variations in interlayer coupling, in contrast to the smooth loops typical of conventional ferroelectrics, such as $Hf_{0.5}Zr_{0.5}O_2$.

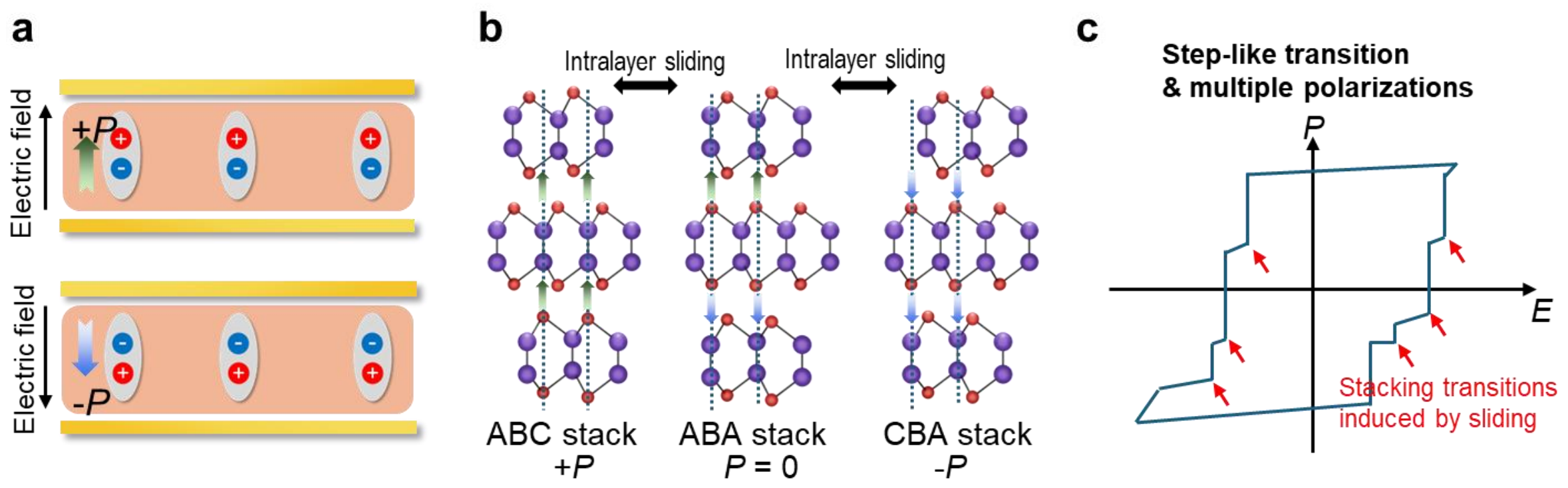

**Box 2. Thickness-dependent band structure evolution of layered InSe.**

**Quantum confinement in layered InSe**

When the motion of an electron is confined in one or more directions, such as in thin films or nanoparticles, the energy becomes quantized into discrete states. This confinement occurs because the electron wave function can no longer spread out within specific boundary conditions. Therefore, in the bulk, electron energy levels are nearly continuous, but they become quantized as thickness is reduced.

As for layered InSe (panel **a**), reducing the number of atomic layers induces an increased band gap due to quantum confinement-induced level separation. In the bulk limit, InSe exhibits a direct band gap of ~1.2 eV at the Γ or B point, with both conduction and valence band edges located at the zone center (panel **b, c**). As thickness decreases, interlayer coupling weakens and confinement effects become more pronounced, particularly in the range of 4–6 layers. Near 5 layers, a transition occurs where the valence band maximum shifts away from the Γ point and drops in energy, marking a crossover from a direct to an indirect band gap. In the monolayer limit, this transition is accompanied by a pronounced distortion in the valence band, leading to symmetric maxima around the Γ point and a camelback-shaped dispersion, with a band gap of ~2.4 eV. This flat-band dispersion enhances the density of states near the valence band edge, introducing a Van Hove singularity and a heavier hole effective mass.

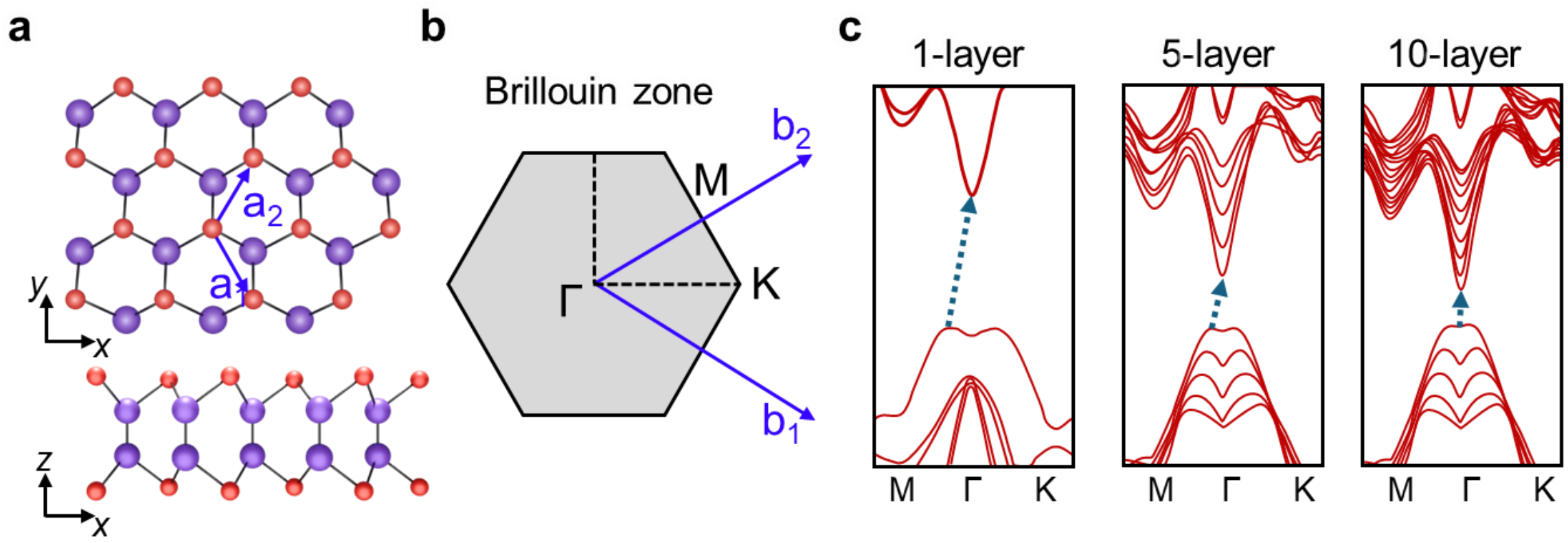

Figures for band structures of InSe are adapted and modified with permission from ref.[157], Springer Nature.

**Supplementary Table 1. Electronic properties of indium selenides.**

| Material[1] | Space Group | Spontaneous Ferroelectricity | Bulk Band Gap (eV) | Monolayer Band Gap (eV) | Effective electron mass ($m_0$)[3] | Thermal velocity (cm/s)[3] | Ref. |
|---|---|---|---|---|---|---|---|
| β-InSe (2H) | $P6_3/mmc$ | No | 1.2 (direct) | | 0.12 | 2.4x$10^7$ | 1,2 |
| ε-InSe (2H) | $P\bar{6}m2$ | No | 1.4 (indirect) | 2.4 (indirect)[2] | 0.12 | 2.4x$10^7$ | 2,3 |
| γ-InSe (3R) | R3m | No | 1.3 (direct) | | 0.12 | 2.4x$10^7$ | 2,4-9 |
| α-$In_2Se_3$ (2H or 3R) | $P6_3/mc$ or R3m | Yes | 1.3 (direct) | 1.5 (direct) | 0.13 | 2.3x$10^7$ | 8,10-12 |
| β-$In_2Se_3$ (1T or 2H or 3R) | $P\bar{3}m1$ or $P\bar{3}m1$ or $R\bar{3}m$ | No | 1.4 (direct) | 1.6 (direct) | 0.12 (theory) | 2.4x$10^7$ | 11-13 |
| β'-$In_2Se_3$ (2H or 3R) | C2/m | Yes | 0.97 (indirect) | 2.5 (indirect) | 0.21 | 1.8x$10^7$ | 14,15 |

[1]Abbreviations used: 1T, trigonal (octahedral coordination); 2H, hexagonal (AB-type stacking); 3R, rhombohedral (ABC-type stacking).
[2]Monolayer InSe exhibits only one possible band gap irrespective of 2H or 3R stacking.
[3]These values correspond to bulk materials rather than few-layer InSe or $In_2Se_3$.

**Supplementary Table 2. Comparisons of ferroelectricity of vdW indium selenides with other 2D ferroelectric materials/systems.**

| Material | OOP/IP direction | Remnant or internal polarization | Coercive field or energy barrier | Curie temp. | Prep. methods |
|---|---|---|---|---|---|
| **α-phase $In_2Se_3$** | OOP and IP (Ref.[16]) | 0.92 μC/cm$^2$ (OOP; Ref.[17]) 0.97 (OOP) and 8.0 μC/cm$^2$ (IP)(DFT; Ref.[18]) | 0.33 V/nm (OOP; Ref.[19]) 300 kV/cm (OOP;Ref.[20]) | 700 K (Ref.[21]) | CVD (Ref.[19,21,22]) ME (Ref.[20,22,23]) |
| **β'-phase $In_2Se_3$** | IP (Ref.[24]) | 0.199 μC/cm$^2$ (DFT; Ref.[24]) | 0.27 eV/unit cell (DFT; Ref.[24]) | 477 K (Ref.[24]) | ME (Ref.[24]) CVD (Ref.[15]) MBE (Ref.[25]) |
| **γ-phase InSe** | OOP (Ref.[26,27]) | 1.0 pC/m (DFT; Ref. [27]) | 0.014 eV/unit cell (DFT; Ref.[27]) | Above RT (Ref. [26,27]) | ME (Ref.[26,27]) |
| **ε-phase InSe** | OOP (Ref.[28]) | 0.04 pC/m (DFT; Ref.[28]) | 0.022 eV/unit cell (DFT; Ref.[28]) | Above RT (Ref.[28]) | ME (Ref.[28]) |
| **$CuInP_2S_6$** | OOP (Ref.[29,30]) | 4 μC/cm$^2$ (200 nm; Ref.[30]) | 300 kV/cm (Ref.[30]) | 320 K (Ref.[29]) | ME (Ref.[30]) |
| **α-phase SnS** | IP (Ref.[31]) | 3 μC/m (Ref.[31]) 17.5 μC/cm$^2$ (Ref. [32]) 2.62 × $10^{-10}$ C/m (DFT; Ref. [33]) | 25 kV/cm (Ref.[31]) 20 kV/cm (Ref.[32]) 10.7 kV/cm (Ref.[34]) | Above RT (Ref.[31,32]) 1200 K (DFT; Ref. [33]) | CVD (Ref.[31,32]) MBE (Ref.[34]) |
| **α-phase SnSe** | IP (Ref.[35]) | 1.5 × $10^{-10}$ C/m (DFT; Ref.[35]) 1.51 × $10^{-10}$ C/m (DFT; Ref. [33]) | - | 380-400 K (Ref.[35]) 326 K (DFT; Ref. [33]) | MBE (Ref.[35]) |
| **α-phase SnTe** | IP (Ref.[36]) | 13-22 μC/cm$^2$ (DFT; Ref.[36]) | - | Above RT for bilayer; 270 K for monolayer (Ref.[36]) | MBE (Ref.[36]) |
| **α-phase GeS** | IP (Ref.[37]) | 5.06 × $10^{-10}$ C/m (DFT; Ref.[33]) | 18.1 kV/cm (Ref.[37]) | 6400 K (DFT; Ref. [33]) | ME (Ref.[37]) |
| **α-phase GeSe** | IP (Ref.[38]) | 3.67 × $10^{-10}$ C/m (DFT; Ref.[33]) | - | 700 K (Ref.[38]) 2300 K (DFT; Ref. [33]) | ME (Ref.[38]) |
| **α-phase Bi** | IP (Ref.[39]) | 0.41 × $10^{-10}$ C/m (DFT; Ref.[39]) | 15.7 mV/Å (DFT; Ref.[39]) | 210 K (Ref.[39]) | MBE (Ref.[39]) |
| **$T_d$-phase $WTe_2$** | OOP (Ref.[40,41]) | 0.19 μC/cm$^2$ (Ref.[40]) $10^4$ e/cm (Ref.[41]) | 0.70 eV/f.u. (Ref.[40]) | 350 K (Ref.[41]) | ME (Ref.[40]) |
| **Bernal-stacked BLG/BN** | OOP (Ref.[42-44]) | 0.9-5.0 pC/m (Ref.[42]) 0.05-0.18 μC/cm$^2$ (Ref.[43]) | 0.2-0.4 V/nm (Ref.[42,43]) | 200 K (Ref.[43]), above RT (Ref. [44]) | ME (Ref.[42-44]) |
| **R-stacked $MoS_2$** | OOP (Ref.[45]) | 0.4 μC/cm$^2$ (Ref.[45]) 0.53 pC/m (Ref.[46]) | 0.036 V/nm (Ref. [45]) | Above RT (Ref.[45,46]) | ME (Ref.[46]) CVD (Ref.[45]) |
| **R-stacked $WSe_2$** | OOP (Ref.[47]) | 1.97 pC/m (Ref.[47]) | 0.3 V/nm (Ref.[47]) | Above RT (Ref.[47]) | ME (Ref.[47]) CVD (Ref.[48]) |

The label "Bernal BLG" refers to Bernal-stacked bilayer graphene sandwiched with hBN layers. “R-stacked” refers to rhombohedral-stacked crystals, which are either naturally grown or CVD-grown ones.

**Supplementary Table 3. Comparisons of precursors for (MO)CVD of 2D TMDs or group-III metal chalcogenides[49-54].**

| | Precursors | Melting temp. (℃) | Decomposition temp. (℃) | Reference |
|---|---|---|---|---|
| **Transition metal precursors** | Mo | 2,623 | | 55,56 |
| | $MoO_3$ | 795 | 780-1200 | 56 |
| | $Mo(CO)_6$ | 150 | 250 | 49,50 |
| | W | 3,422 | | 55 |
| | $WO_3$ | 1,473 | 750-900 | 55,57 |
| | $W(CO)_6$ | 170 | 350 | 50 |
| **Group-III metal precursors** | In | 157 | | 58 |
| | InI | 365 | | 59 |
| | $In(CH_3)_3$ | 88 | 101-320 | 54 |
| | $In_2O_3$ | 1,910 | 2,000 | 60 |
| | Ga | 29.8 | | 58 |
| | $Ga(CH_3)_3$ | -15.7 | 150-475 | 53 |
| | $Ga_2O_3$ | 1793 | 1800 | 61 |
| **Chalcogen precursors** | S | 120 | | 55 |
| | Se | 221 | | 55 |
| | $(CH_3)_2S$ | 37.33* | 400-675 | 51 |
| | $(C_2H_5)_2S$ | 92.1* | 500 | 49 |
| | $(CH_3)_2Se$ | 49.8* | 400 | 62 |

Mark * indicates boiling temperature instead of melting temperature.

**Supplementary Table 4. Native point defects and extrinsic dopants in indium selenides.**

| Types | Materials | Defect types or dopants | Doping type | Notes | Reference |
|---|---|---|---|---|---|
| **Native point defects** | InSe | Selenium vacancy ($V_{Se}$) | Deep-level trap | Acts as a recombination center, reducing carrier lifetime | 63 |
| | | Indium interstitial ($In_i$) | Shallow donor | n-type contribution | 63 |
| | | Indium vacancy ($V_{In}$) | Shallow acceptor | p-type contribution | 63 |
| | | In-in-Se antisite ($In_{Se}$) | Not contribute to doping | Energetically expensive and rare | 63 |
| | | Se-in-In antisite ($Se_{In}$) | Both a donor and an acceptor | Dominant defect in Se-rich materials | 63 |
| | **α-$In_2Se_3$** | Indium vacancy ($V_{In}$) | Shallow acceptor- | p-type contribution | 64 |
| | | Selenium vacancy ($V_{Se}$) | Shallow donor- | n-type contribution | 64 |
| **Extrinsic dopants** | InSe | **$Sn_{In}$** | Shallow donor | Enhanced electron mobility and phototransistor responsivity | 65 |
| | | **$As_{Se}$** | Shallow acceptor | Introduction of acceptor levels | 66 |
| | | **$Y_{In}$** | Semimetal | Semiconductor-to-semimetal conversion | 67 |
| | **α-$In_2Se_3$** | **$I_{Se}$** | Shallow donor- | Increased electron concentration | 68 |
| | | **$Cd_{In}$** | | Decreased out-of-plane conductivity and increase in conductivity anisotropy. | 68 |
| | | **$Cu_{In}$** | | | 68 |

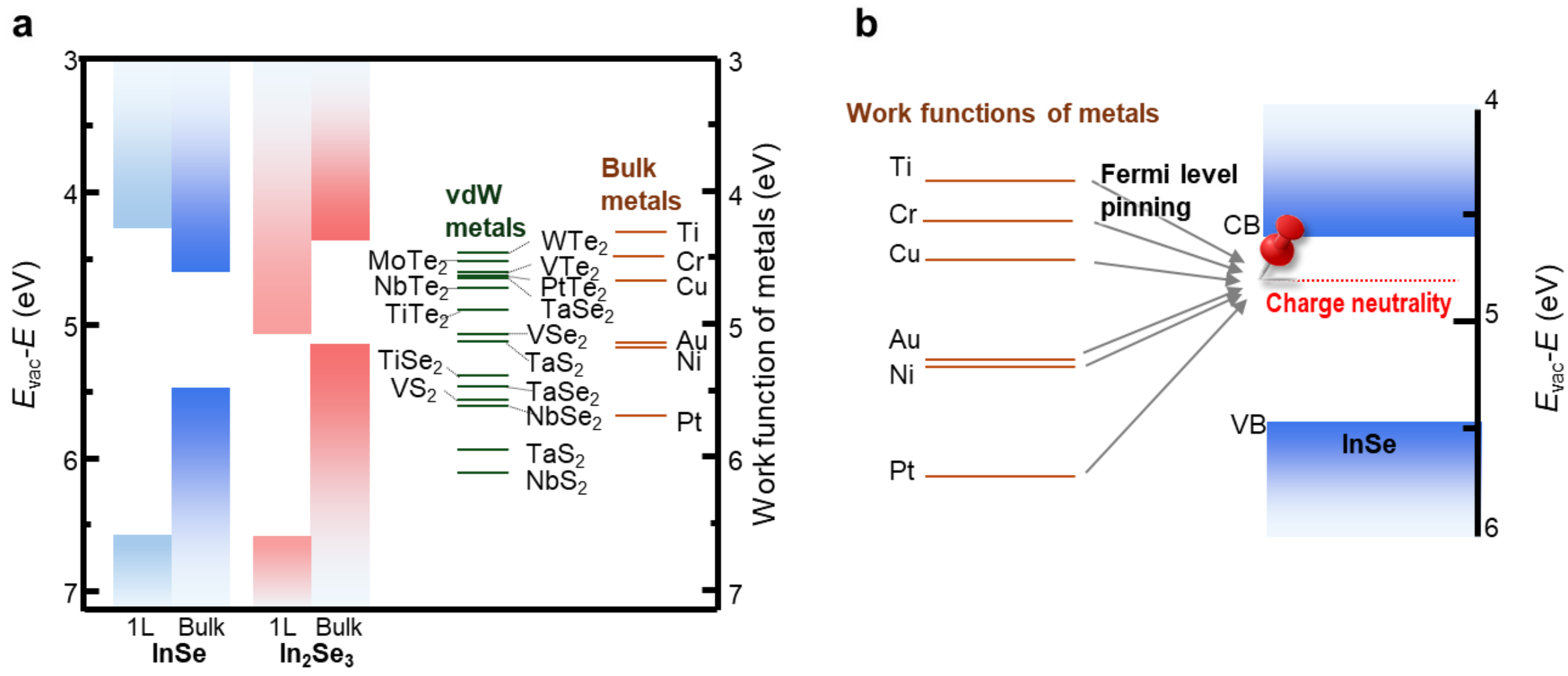

**Supplementary Fig. 1. Fermi-level pinning in 2D electronic devices based on indium selenides.** (**a**) Schematic energy-band diagrams illustrating the band structures of monolayer (labeled as 1L) and multilayer (labeled as Bulk) InSe (blue) and α-phase $In_2Se_3$ (red) on the left y-axis, together with the work functions of various metals on the right y-axis. (**b**) Illustration of Fermi-level pinning caused by interface states, dangling bonds, and defects at the metal–semiconductor junction of 2D InSe (blue, right) contacted with 3D metals (brown, left). These states localize the Fermi level near a charge-neutrality level (red dashed line) within the bandgap, thereby reducing the tunability of the Schottky barrier and limiting modulation of contact resistance through metal selection. See Supplementary Note 1 for more information.

## Supplementary Note 1. Fermi-level pinning in 2D indium selenides.

**Fermi-level pinning at metal-semiconductor interfaces**

When a metal contacts a semiconductor, the ideal Schottky barrier height ($\Phi_B$) is determined by the difference between the metal's work function ($\Phi_M$) and the semiconductor's electron affinity ($\chi$) as; $\Phi_B = \Phi_M - \chi$ (for electrons) (see band structures of monolayer and multilayer InSe and α-$In_2Se_3$ and work function of various metals in **Supplementary Fig. 1a**). However, in practice, the Fermi level often becomes "pinned" at a specific energy level, largely independent of the metal's work function. This phenomenon, known as Fermi level pinning, is caused by interface states such as metal-induced gap states (MIGS), defects, or dangling bonds that form at the junction. These states trap charge carriers and localize the Fermi level near specific energy positions (i.e., charge neutrality) within the bandgap, suppressing the tunability of the Schottky barrier and restricting the ability to modulate contact resistance through careful metal selection (**Supplementary Fig. 1b**).

**Contact engineering in indium selenides**

In 2D layered semiconductors, Fermi-level pinning is often more severe because their electronic properties are highly sensitive to interfacial effects and metal deposition conditions. For example, 2D InSe FETs with In metal contacts exhibits Fermi-level pinning effects due to chemical disorder and MIGS[69], that induces higher Schottky barrier height than the ideal case. Besides, Se vacancies can act as trap states that shift the Fermi level and increase the Schottky barrier[70]. Various approaches with a contact metal with van der Waals contact interface have been studied to above this issue for indium selenides devices[71]. Inserting an oxidized monolayer at the contact interface is also effective to suppress metal- and disorder-induced gap states[72].

## Supplementary References